# Letter to the Editor
# A cosmological test with compact radio sources

R. Kayser

Hamburger Sternwarte, Gojenbergsweg 112, D-21029 Hamburg-Bergedorf, Germany

**Abstract.** The recent proposal by Kellermann (1993) to use the angular size/redshift relation of compact radio sources to obtain information on the value of cosmological parameters has provoked considerable discussion in the astronomical community. I present here a careful re-analysis of the angular size/redshift data of radio jets used by Kellermann (1993), using a Kolmogorov-Smirnov test to check the independence of the distribution of linear jet sizes from redshift under the assumption that no evolution is present in the sources. The test takes into account biasing of the sample due to the limited resolution.

While the data are compatible with an Einstein-de Sitter Universe ($P = 0.2$), they are not sufficient to rule out a significant part of the $(\Omega_0, \lambda_0)$-plane.

## 1. Introduction

Recently Kellermann (1993) made the interesting proposal to use the angular size/redshift relation of compact radio sources to obtain information on the cosmological deceleration parameter $q_0$, which in the case of a vanishing cosmological constant equals $\Omega_0/2$.

While earlier attempts to use the angular size/redshift relation for high redshift optical or radio sources have been faced with severe problems due to the difficulty of precisely defining and measuring angular diameters, as well as excluding evolutionary effects, compact radio sources seemingly offer a way around these obstacles.

Kellermann (1993) concluded that the data are consistent with an Einstein-de Sitter Universe and that they provide direct evidence that the density parameter does not differ significantly from unity. However, this conclusion is based only on visual comparison of the binned data with the theoretical angular size/redshift relation for models with a vanishing cosmological constant.

In this letter a careful re-analysis of the original unbinned data is presented, without a priori restrictions on the cosmological model, using a Kolmogorov-Smirnov test and taking into account biasing due to the limited angular resolution in the sample.

## 2. The sample

For his analysis Kellermann (1993) used a sample of 82 compact radio sources taken from the existing literature which met his selection criteria discussed below. A publication of the data of the complete sample, including references, is in preparation (Kellermann 1994). I will here give a short overview of the selection criteria as well as of the definition and measurement of the angular size. For a complete description see Kellermann (1993).

In order to confine the sample to a homogeneous group with similar physical properties, only sources with a luminosity greater than $10^{24}$ W Hz$^{-1}$ (at $\lambda = 6$ cm) have been used. This is roughly the luminosity which separates radio-loud from radio-quiet quasars. Furthermore, Kellermann (1993) excluded all known BL Lac objects, compact central components of extended double-lobed quasars, and peaked spectrum sources.

In addition, I excluded from the sample not only unresolved sources but also sources with redshifts below $z = 0.03$, for which the Doppler redshift due to peculiar motion may not be negligible. This reduced the number of objects from 82 to 76.

Kellermann defined the angular size $\theta$ of a source as the distance between the core and the most distant component whose peak brightness exceeds 2% of the core brightness. This definition of the angular size should keep any effect of observing at a fixed frequency small. All sizes were determined from the published contour diagrams based on VLBI observations at 6 cm, a resolution $\theta_0$ of about 1 milliarcsecond, and a dynamic range of at least 1:100.

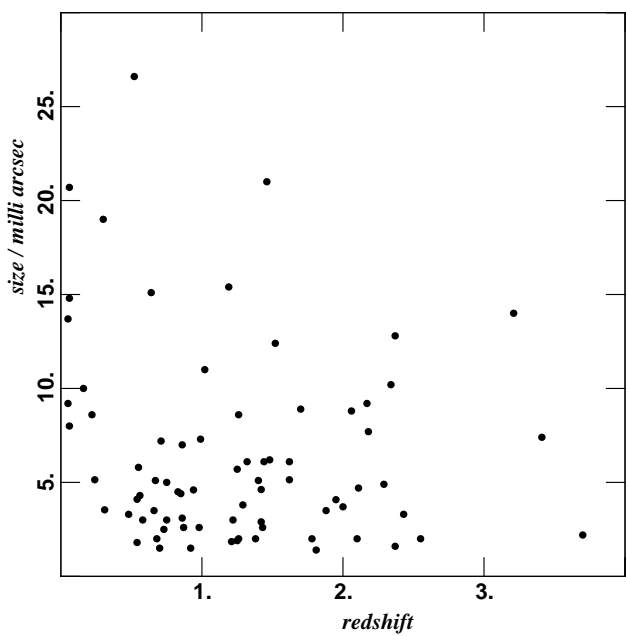

**Fig. 1.** Angular size/redshift diagram for the complete sample of compact radio sources used in this analysis.

## 3. The statistical method

My aim is, within the framework of standard Friedmann-Lemaître cosmology, to obtain information on the cosmological parameters $\Omega_0$ (density parameter) and $\lambda_0$ (normalised cosmological constant, $\lambda_0 = \Lambda/(3H_0^2)$) from the distribution of angular jet sizes $\theta$ in dependence on the redshift $z$.

Kellermann (1993) collected the data into redshift bins, calculated the mean angular size for each bin, and than compared visually the binned data to the expected angular size/redshift relation for $\lambda_0 = 0$ models.

While this approach makes sense to get a first overview over the data and to demonstrate in an illustrative way how the proposed test works, it is clear that a more sound statistical test is needed to obtain a 'best fit' world model or to reject world models incompatible with the data. In addition, no a priori restriction (like setting $\lambda_0 = 0$) should be made for the cosmological model.

Binning the data introduces several problems into the analysis (see, e.g., Press et al. 1986). Since each bin represents a subsample of the data by just one quantity (the mean), information on the inner distribution of the data is lost. The mean of the data is in most cases (including this one) not representative of the value of the considered function at the centre of the bin. Finally, binning introduces free parameters (size and positions of the bins) into the analysis, on which the results obviously depend in some difficult to control way. The latter is especially true if the number of bins and the number of objects per bin is small.

data without binning for the analysis. Fig. 1 clearly shows that the distribution of the angular size $\theta$ is not a normal (Gaussian) one. There is some concentration towards small $\theta$ and a significant tail at high $\theta$. Thus, a standard $\chi^2$ approach cannot be used to fit a $\theta(z)$ relation to the data.

Assuming that the radio jets show no evolution with redshift is equivalent to assuming that the distribution of linear jet sizes $L'$ is independent of the redshift. It seems natural to test this assumption by means of a Kolmogorov-Smirnov test. To do this, I split the sample into a low redshift ($z < z_{\rm cut}$) and a high redshift ($z > z_{\rm cut}$) subsample, and check whether the distribution of the linear sizes (for the considered world model) is the same for both subsamples.

The linear size of the jets is obtained from

$$L' = D(H_0, \Omega_0, \lambda_0; z)\theta \tag{1}$$

where $D = rR$ is the angular size distance (see, e.g., Berry 1989), which depends on the Hubble parameter $H_0$, the world model $(\Omega_0, \lambda_0)$ and the redshift $z$ of the source. Since I am only interested in the distribution of the jet sizes and not in their absolute values, I use only dimensionless linear sizes defined as

$$L = \frac{H_0}{c} L' = \frac{H_0}{c} D\,\theta \tag{2}$$

which do not depend on $H_0$ since $D \propto H_0^{-1}$.

While the sample is (hopefully) complete with respect to $\theta$ down to the resolution limit $\theta_0$, it is not complete with respect to the linear sizes $L$. Sources with linear sizes

$$L < D(\Omega_0, \lambda_0; z)\,\theta_0 \tag{3}$$

are not resolved, a limit obviously depending on the redshift and the world model. For each world model the angular size distance has a certain maximum value $D_{\rm max}$ for the considered sample. I denote

$$L_0 = D_{\rm max}(\Omega_0, \lambda_0)\,\theta_0 \tag{4}$$

as the linear resolution limit for the world model $(\Omega_0, \lambda_0)$. It is clear that the sample is incomplete for $L < L_0$. Sources smaller than the linear resolution limit are not resolvable at all redshifts.

For each considered world model I must therefore first calculate $D_{\rm max}$ and $L_0$, and then exclude all objects with $L < L_0$ to avoid biasing due to the limited resolution.

The Kolmogorov-Smirnov test applied to the two subsamples gives the probability $P$ that the subsamples are drawn from the same parent distribution. This is always an upper limit to the probability that the distribution of $L$ is independent of the redshift.

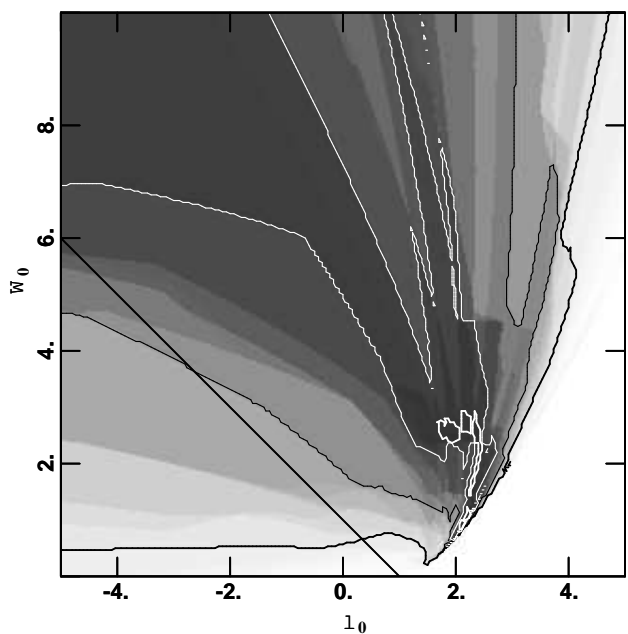

**Fig. 2.** Probability $P(\Omega_0, \lambda_0)$ that the distribution of linear jet sizes is independent of the redshift. White corresponds to $P = 0$. The thick white line, the thin white line, the thin black line and the thick black line correspond to $P = 0.99, 0.9, 0.5$, and $0.1$, respectively. The straight black line marks flat universes ($k = 0$).

## 4. Results and discussion

I have tested $300 \times 300$ Friedmann-Lemaître models in the range $\Omega_0 = 0 \cdots 10$ and $\lambda_0 = -5 \cdots +5$, using the method described above. In order to keep the number of objects in both subsamples and for all tested world models above 20, which is necessary to obtain reliable results from the Kolmogorov-Smirnov test (Press et al. 1986), the sample is split at $z_{\rm cut} = 0.75$. In Fig. 2 the resulting probability distribution is shown.

The highest probability ($P > 0.99$) is found for $\Omega_0 = 1.0 \cdots 3.0$ and $\lambda_0 = 1.5 \cdots 2.5$. This should, however, not be interpreted as a "best fit", since high probabilities in KS-tests usually occur just by chance, while the true value of the KS-test is the rejection of models with low probabilities ($P < 0.01$ is a conservative significance level).

Marginal evidence ($P < 0.1$) can be found for the existence of (non-baryonic) dark matter, since a minimum value of $\Omega_0 = 0.2$ is found for $P > 0.1$, which is larger than the current limit for baryonic matter as obtained from nucleosynthesis (see, e.g., Krauss 1994). If one assumes $\lambda_0 = 0$, a minimum of $\Omega_0 = 0.53$ is required for $P > 0.1$. If one assumes a flat universe ($k = 0$, i.e. $\Omega_0 + \lambda_0 = 1$), $\Omega_0 = 0.67$ is required for $P > 0.1$.

The data are compatible with the Einstein-de Sitter case $\Omega_0 = 1$ and $\lambda_0 = 0$ at a moderate probability of $P = 0.2$.

icant part of the $(\Omega_0, \lambda_0)$ plane at a sufficient significance level ($P < 0.01$). Note that the white area ($P = 0$) in Fig. 2 corresponds to so-called bounce models (Stabell & Refsdal 1966) which are rejected by the mere existence of redshifts up to $z = 4$.

The different result of Kellermann (1993) is mainly an artefact of the binning scheme and of the a priori restriction to $\lambda_0 = 0$ models. Due to the large intrinsic scatter of the sizes of compact radio sources, their angular size/redshift relation does not provide us with any evidence that the cosmological constant is close to zero or that the density parameter is close to unity.

*Acknowledgements.* I am especially grateful to Ken Kellermann for providing me with his original radio jet data, as well as for interesting discussions. Further thanks to A. Dent and P. Helbig for reading the manuscript and helpful discussions.